\documentclass[fleqn,twoside]{article}
\usepackage{espcrc2}

\usepackage{graphicx}

\newcommand{\be}{\begin{equation}}
\newcommand{\ee}{\end{equation}}
\newcommand{\ben}{\begin{equation*}}
\newcommand{\een}{\end{equation*}}
\newcommand{\beq}{\begin{eqnarray}}
\newcommand{\eeq}{\end{eqnarray}}

\newcommand{\AmS}{{\protect\the\textfont2
  A\kern-.1667em\lower.5ex\hbox{M}\kern-.125emS}}

\title{Calculation of the N to $\Delta$ 
electromagnetic transition matrix element~\thanks{Talk presented by A.~Tsapalis}}		

\author{C.~Alexandrou\address{Department of Physics, 
University of Cyprus, CY-1678 Nicosia, Cyprus},
Ph.\ de Forcrand\address{ETH-Z\"urich, CH-8093 Z\"urich and CERN Theory Division, CH-1211 Geneva 23, Switzerland},
Th.~Lippert\address[wuppertal]{Department of Physics, University of
Wuppertal, D-42097 Wuppertal, Germany},
H.~Neff\thanks{Acknowledges funding from 
 the European network ESOP (HPRN-CT-2000-00130) and  the
University of Cyprus.}\address{Institute of Accelerating Systems and Applications and
Department of Physics, University of Athens, Athens, Greece},
J.~W.~Negele\address[MIT]{Center for Theoretical Physics, Laboratory for
Nuclear Science and Department of Physics, Massachusetts Institute of
Technology, Cambridge, Massachusetts 02139 U.S.A},
K.~Schilling\addressmark[wuppertal],
W. Schroers\addressmark[MIT]
and
A.~Tsapalis\addressmark[wuppertal]
}
       
\begin{document}

\begin{abstract}
We present  results on  the ratio of electric quadrupole 
to magnetic dipole amplitudes, $R_{EM}={\cal G}_{E2}/{\cal G}_{M1}$,
  for the transition $\gamma N\rightarrow \Delta$
from lattice QCD.
We consider both the quenched and the 2-flavor theory.
\vspace{1pc}
\end{abstract}

\maketitle

\vspace*{-1.8cm}

\section{INTRODUCTION}

Accurate experimental measurements of transition matrix elements $N \rightarrow
\Delta$~\cite{Athens,Joo} give strong support for a deformed nucleon and/or $\Delta$.
On the lattice, hadron wave functions~\cite{wfs}
can  provide information on the deformation of particles of spin higher than 1/2. 
For a spin 1/2 particle however, the quadrupole moment vanishes,
and this is the reason why
one turns to measurements of quadrupole strength
in the  $\gamma N \> \rightarrow \> \Delta$ transition
as in experimental studies.
State-of-the-art lattice QCD calculations can yield accurate 
results on these matrix elements and provide direct comparison with experiment.
Spin-parity selection rules allow a magnetic dipole, M1, an electric
quadrupole, E2, or a Coulomb quadrupole, C2, amplitude.
If both the nucleon and the $\Delta$
are spherical then E2 and C2 are
expected to be zero. Although M1 is indeed the dominant amplitude
there is mounting experimental evidence for a range of momenta transfer
that E2 and C2 are
 non-zero~\cite{Athens,Joo}.
The physical origin of non-zero  E2 and C2 amplitudes is
attributed to different mechanisms in the various models.
In quark models the deformation is due to the colour-magnetic tensor
force.
In ``cloudy'' baryon models it is due to meson exchange
currents.
By comparing quenched and unquenched results we aim at an understanding
of the origin of a non-zero deformation.

A previous lattice QCD study \cite{Leinweber}
of the $N \> \rightarrow \> \Delta$  transition
 with a limited number of
quenched configurations yielded an inconclusive result for the ratio,
 $R_{EM}$,
 of E2 to M1 amplitudes, 
 since
a zero value could not be statistically excluded.  
In this work we improve by using smearing techniques to have cleaner plateaus,
a  larger lattice and more configurations. 
Moreover, we present first full QCD results, based on the SESAM lattices
~\cite{SESAM}.

\vspace*{-0.3cm}

\section{LATTICE MATRIX ELEMENTS}
The current matrix element for $N \> \rightarrow \> \Delta$
transitions with on-shell nucleon and $\Delta$ states and real or
virtual photons has the form \cite{Jones73}

\vspace*{-0.5cm}

\beq
 \langle \; \Delta (p',s') \; |& J_\mu &| \; N (p,s) \rangle = \nonumber \\
 &\>& \hspace*{-2.0cm}i \sqrt{\frac{2}{3}} \biggl(\frac{M_{\Delta}\; M_N}{E'_{\Delta}\;E_N}\biggr)^{1/2}
 \bar{u}_\tau (p',s') {\cal O}^{\tau \mu} u(p,s) \;.
\eeq
The operator 
${\cal O}^{\tau \mu}$  can be decomposed as
\be
  {\cal G}_{M1}(q^2) K^{\tau \mu}_{M1} 
+{\cal G}_{E2}(q^2) K^{\tau \mu}_{E2} 
+{\cal G}_{C2}(q^2) K^{\tau \mu}_{C2} \;,
\ee
where the magnetic dipole, ${\cal G}_{M1}$, electric quadrupole, 
${\cal G}_{E2}$,
 and scalar
quadrupole, ${\cal G}_{C2}$, form factors depend on the momentum
transfer $q^2 = (p'-p)^2$. The kinematical functions 
$K^{\tau \mu}$ depend on $p,\>p',\>M_N$ and $M_\Delta$ and their expressions
 are given in \cite{Jones73}.
Following \cite{Leinweber} we calculate  the three-point correlation
functions 
\beq
\langle G^{\Delta j^\mu N}_{\sigma} 
(t_2, t_1 ; \vec{p}^{\;\prime}, \vec{p}; \Gamma) \rangle &=& \nonumber \\
&\>& \hspace*{-4.5cm}\sum_{\vec{x}_2, \;\vec{x}_1}
\exp(-i \vec{p}^{\;\prime} \cdot \vec{x}_2 )  
\exp(+i (\vec{p}^{\;\prime} -\vec{p}) \cdot \vec{x}_1 ) \;  \nonumber \\
&\>&\hspace*{-4.2cm}\Gamma^{\beta \alpha}
\langle \;\Omega \; | \; T\left[\chi^{\alpha}_{\sigma}(\vec{x}_2,t_2) 
j^{\mu}(\vec{x}_1,t_1) \bar{\chi}^{\beta} (\vec{0},0) \right]
\; | \;\Omega \;\rangle \quad. 
\eeq
and 
 $\langle G^{N j^\mu \Delta}_{\sigma} 
(t_2, t_1 ; \vec{p}^{\;\prime}, \vec{p}; \Gamma) \rangle $.
For the spin-$\frac{1}{2}$,~$\chi ^p (\vec{x},t)$, and -$\frac{3}{2}$, 
$\chi ^{\Delta^{+}}_\sigma  (\vec{x},t)$, interpolating fields 
and projection matrices $\Gamma$ we use the expressions  given in ref.~\cite{Leinweber}.

For large Euclidean time separations 
$t_2 -t_1 \gg 1$ and $t_1 \gg 1$,
the time dependence and field normalization constants are cancelled
 in the following ratio~\cite{Leinweber}
\beq
R_\sigma (t_2, t_1; \vec{p}^{\; \prime}, \vec{p}\; ; \Gamma ; \mu)& =&\nonumber \\
&\>&\hspace*{-4.7cm}\large{
\left[ \frac{
\langle G^{\Delta j^\mu N}_{\sigma} (t_2, t_1 ; \vec{p}^{\;\prime}, \vec{p};
\Gamma ) \rangle \;
\langle G^{N j^\mu \Delta}_{\sigma} (t_2, t_1 ; -\vec{p}, -\vec{p}^{\;\prime};
\Gamma^\dagger ) \rangle }
{
\langle -g_{ij} G^{\Delta \Delta}_{ij}(t_2,\vec{p}^{\; \prime};
\Gamma_4) \rangle \;
\langle G^{NN} (t_2, -\vec{p} ; \Gamma_4) \rangle } \right]^{1/2}
} \nonumber \\
&\>&\hspace*{-4.5cm}\sim \biggl( \frac{E_N+M_N}{2E_N} \biggr)^{1/2} \biggl( \frac{E'_\Delta+M_N}{2E'_\Delta} \biggr)^{1/2} \bar{R}_{\sigma}(\vec{p}^{\; \prime}, \vec{p}\; ; \Gamma ; \mu)
\label{R-ratio}
\eeq
\normalsize 
We use the lattice conserved   electromagnetic current,   $j^\mu (x)$,
symmetrized on site $x$ by taking
$
j^\mu (x) \rightarrow \left[ j^\mu (x) + j^\mu (x - \hat \mu) \right]/ 2
$.

In the rest frame of the $\Delta$ we take $ -\vec{p}=\vec{q}=(q,0,0)$ and
$ \vec{p}^{\;\prime}=(0,0,0)$.
The Sachs form factors can be extracted 
from the  plateau values of $ \bar{R}_{\sigma}(\vec{p}^{\; \prime}, \vec{p}\; ; \Gamma ; \mu)$ for specific combinations of matrices $\Gamma$ and 
$\sigma$ as in ref.~\cite{Leinweber}.
For example, $R_{EM}$ is given by

\be
R_{EM}=-\large{
\frac{1}{3}\frac{\lbrace \bar{R}_3 (\vec{q}, 0\; ; \Gamma_1 ; 3) +
\bar{R}_1 (\vec{q}, 0\; ; \Gamma_3 ; 3) \rbrace}
{
\lbrace \bar{R}_3 (\vec{q}, 0\; ; \Gamma_1 ; 3) -
\bar{R}_1 (\vec{q}, 0\; ; \Gamma_3 ; 3) \rbrace
}} .
\ee
\normalsize
\begin{figure}[h]
\vspace*{-0.5cm}
\mbox{\includegraphics[height=4.5cm,width=7cm]{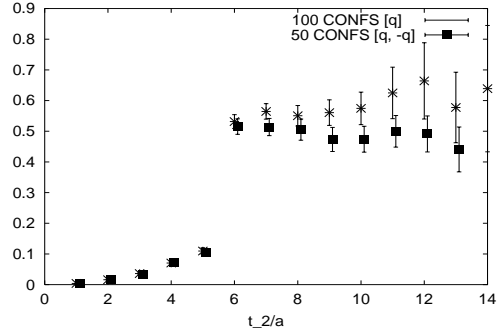}}
\vspace*{-1cm}
\caption{Noise reduction in  $\bar{R}_3 (\vec{q}, 0\; ; \Gamma_1 ; 3)$
   by equal $U, \; U^*$ weighting in the action 
for 100 SESAM configurations at $\kappa=0.1570$. $t_1/a=6$ and $a$ is  the  lattice spacing.}
\label{fig:noisereduction}
\vspace*{-0.8cm}
\end{figure}

Smearing is essential for filtering  the ground
state before the signal from the time correlators 
is lost in the noisy large time limit.
We use the gauge invariant Wuppertal smearing 
in order to increase the overlap with the baryon states.
 Quark
propagators with a photon insertion are computed with the sequential
source technique. We also implement the equal re-weighting of 
$\lbrace U \rbrace $ and $\lbrace U^* \rbrace $ gauge configurations  
in the lattice action~\cite{Draper}. Fig~1 shows that 
fluctuations  are substantially
reduced at large time separations resulting in better plateaus for 
$ \bar{R}_{\sigma}(\vec{p}^{\; \prime}, \vec{p}\; ; \Gamma ; \mu)$.  
 This greatly overbalances the fact that  
 an additional sequential
propagator with momentum \\
(-q,0,0) is required.

\section{RESULTS}
\begin{figure}[h]
\vspace*{-1.0cm}
\mbox{\includegraphics[height=5cm,width=7cm]{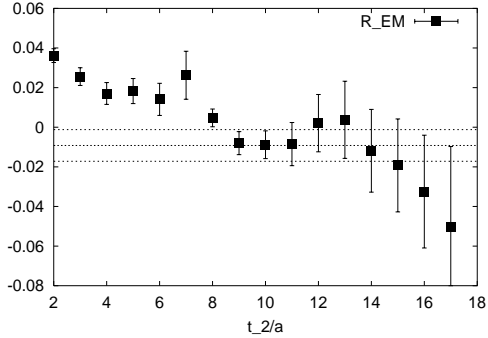}}
\vspace*{-1cm}
\caption{$R_{EM}$ at $\kappa=0.1558$ for 100 quenched confs.
The photon is injected~at~$t_1/a=8$.}
\label{fig:emr_k0.1558_quenched}
\vspace*{-0.8cm}
\end{figure}

For the quenched analysis we used 100 gauge configurations
on a lattice of size $32^3\times 64$ at $\beta=6.0$. We take 
$u$ and $d$ quarks of the same mass.
The lowest momentum transfer on this lattice is $q^2\approx 0.14 \;{\rm GeV}^2$,
close to the value of  $q^2=0.126 \;{\rm GeV}^2$ where
an accurate measurement of $R_{EM}=(-2.1 \pm 0.2_{stat} 
\pm 2.0_{mod})\%$ was recently 
 obtained~\cite{Athens}. Note that the statistical error is an order of 
magnitude smaller than the `model' estimated error involved
in the extraction of this ratio. A reliable result from lattice QCD
is thus a valuable input.  
A photon is injected at $t_1/a = 8$, allowing enough time for filtering out
excited baryonic states.
 Jackknife averages and $\chi^2$
fits are performed on the plateaus.
$R_{EM}$ at $\kappa = 0.1558$ is shown in 
Fig.~\ref{fig:emr_k0.1558_quenched}.
The best fit to the plateau yields
$R_{EM}= (-0.9 \pm  0.8) \%$ with $\chi^2 /{\rm d.o.f} \sim 0.4$. 

If deformation is due to the pion cloud, unquenching should make
the ratio $R_{EM}$ more negative. 
We analysed 100 SESAM configurations for a lattice of size $16^3 \times 32$ 
at $\beta=5.6$
at four different values of sea quark mass. Preliminary results
are given in the Table.
The analysis was done at the  lowest possible momentum
transfer for this lattice, $q^2 \sim 0.53 {\rm GeV}^2$
(taking $a^{-1}=1.85$~GeV from the chiral extrapolation of the nucleon mass). 

\begin{table}[h]
\vspace*{-0.7cm}
\begin{tabular}{|cccc|} \hline
$\kappa_{sea}$  & $m_\pi/m_\rho$ & $R_{EM} (\%)$ & $R_{SM} (\%)$ 
\\  \hline     
$0.1560$  & $0.83$   &  $-2.24 \pm 0.46$ &    \\
$0.1565$  & $0.81$   &  $-2.25 \pm 0.55$ &  \\
$0.1570$  & $0.76$   &  $-3.40 \pm 0.61$ &  $-3.2 \pm 2.1$ \\
$0.1575$  & $0.69$   &  $-2.98 \pm 0.90$ & \\ 
\hline
\end{tabular}
\vspace*{-0.8cm}
\end{table}

\begin{figure}[h]
\vspace*{-1.cm}
\mbox{\includegraphics[height=5cm,width=7cm]{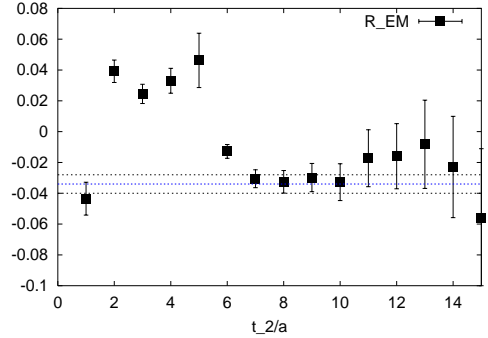}}
\vspace*{-1cm}
\caption{$R_{EM}$ ratio for 100 SESAM confs. at  $\kappa=0.1570$.
The photon is injected at $t_1/a=6$.}
\label{fig:emr_k0.1570_sesam}
\vspace*{-0.8cm}
\end{figure}

\begin{figure}[h]
\vspace*{-0.5cm}
\mbox{\includegraphics[height=5cm,width=7cm]{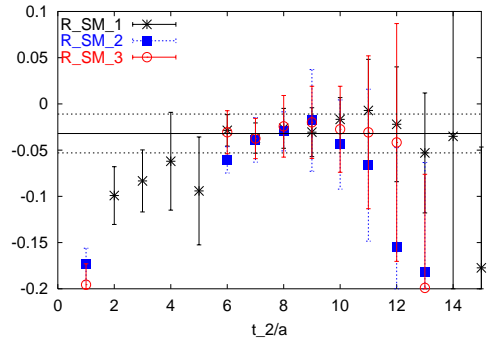}}
\vspace*{-0.9cm}
\caption{$R_{SM}$ plateaus for 100 $\kappa=0.1570$ SESAM lattices. The three equivalent definitions are consistent within the errors. $t_1/a=6$.}
\label{fig:cmr_k0.1570_sesam}
\vspace*{-0.7cm}
\end{figure}

The measured $R_{EM}$ as a function of the sink time-slice $t_2$
is shown in Fig.~\ref{fig:emr_k0.1570_sesam} for $\kappa_{sea}=0.1570$. 
We obtain a very clear plateau at a negative value.
Moreover, this value, in the  ($-2 $ to $-3$)\% range,
is in agreement with the recently measured experimental value of $-1.6 \pm 0.4
\pm 0.4$ at a similar momentum transfer~\cite{Joo}. 
Considering that the experimental  value of $R_{EM}$ changes little
from $q^2=0.126$~GeV$^2$ to  $q^2=0.52$~GeV$^2$,
we expect that the main difference between our quenched and unquenched
results comes from dynamical quark contributions  
pointing to  pionic cloud contributions to the deformation.
For the unquenched case we have also measured the ratio of C2 to M1 amplitudes,
 $R_{SM}$, which involves the time component of the electromagnetic
current.
In Fig.~\ref{fig:cmr_k0.1570_sesam} we show this ratio
as a function of time. The C2 amplitude is extracted
from  the three combinations  
$\bar{R}_1(\vec{q},0;-i\Gamma_1;4)$, $\bar{R}_2(\vec{q},0;i\Gamma_2;4)$
and $\bar{R}_3(\vec{q},0;i\Gamma_3;4)$. As it can be seen all three yield
consistent results. Fitting to the common plateau we obtain 
a negative non-zero result indicating baryon deformation. 

In summary, using state-of-the-art lattice techniques 
the phenomenologically 
important $R_{EM}$ and $R_{SM}$ values can be  extracted for various $q^2$ values.
Unquenching drives both ratios  more negative and yields results
that are consistent with 
experimental measurements.


\vspace*{-0.3cm}

\begin{thebibliography}{99}
\bibitem{Athens} C.Mertz {\it et al.}, Phys. Rev. Lett. {\bf 86} (2001) 2963.
\bibitem{Joo} K. Joo {\it et al.}, 
Phys.\ Rev.\ Lett.\ {\bf 88} (2002) 122001.
\bibitem{wfs} C. Alexandrou, Ph. de Forcrand and A. Tsapalis,
hep-lat/0206026; these proceedings. 
\bibitem{Leinweber} D. B. Leinweber, T. Draper and R. M. Woloshyn,
Phys. Rev. D {\bf 48}, (1993) 2230.
\bibitem{SESAM} N. Eicker  {\it et al.}, Phys. Rev. D {\bf 59} (1999) 
014509.
\bibitem{Jones73} H. F. Jones and M.C. Scadron, Ann. Phys. (N.Y.) {\bf81},
1 (1973)
\bibitem{Draper} T. Draper, W. Wilcox, R. M. Woloshyn and K. F. Liu,
Nucl. Phys. B {\bf 318}, (1989) 319; Nucl. Phys. B (Proc. Suppl.){\bf 9}, (1989) 175.   

\end{thebibliography}
\end{document}